\documentclass{article}
\usepackage{spconf,amsmath,graphicx}
\usepackage{etoolbox}
\usepackage{booktabs}
\usepackage{url}
\usepackage{tabularx}
\usepackage[hidelinks]{hyperref}
\usepackage{mathtools}
\usepackage{xcolor}
\usepackage{cite}
\usepackage{multirow}
\usepackage{subcaption}

\patchcmd{\thebibliography}
  {\settowidth}
  {\setlength{\itemsep}{0pt plus 0.1pt}\settowidth}
  {}{}

\DeclareRobustCommand{\sbseries}{\fontseries{sb}\selectfont}
\DeclareTextFontCommand{\textsb}{\sbseries}

\title{Efficient Speech Quality Assessment using Self-supervised Framewise Embeddings}
\name{
    Karl El Hajal$^{1,2}$*\thanks{*KEH (karlhajal@gmail.com) performed this work as an intern at Logitech.},
    Zihan Wu$^{1,2}$,
    Neil Scheidwasser-Clow$^{3}$,
    Gasser Elbanna$^{1,4}$,
    Milos Cernak$^{2}$
}
\address{
    $^{1}$\'Ecole Polytechnique F\'ed\'erale de Lausanne (EPFL), Lausanne, Switzerland,\\
    $^{2}$Logitech Europe, Lausanne, Switzerland,\\
    $^{3}$University of Copenhagen, Copenhagen, Denmark,\\
    $^{4}$McGovern Institute for Brain Research MIT, Cambridge, MA, USA
}
\begin{document}
\ninept
\maketitle
\begin{abstract}

    Automatic speech quality assessment is essential for audio researchers, developers, speech and language pathologists, and system quality engineers. The current state-of-the-art systems are based on framewise speech features (hand-engineered or learnable) combined with time dependency modeling. This paper proposes an efficient system with results comparable to the best performing model in the ConferencingSpeech 2022 challenge. Our proposed system is characterized by a smaller number of parameters (40-60x), fewer FLOPS (100x), lower memory consumption (10-15x), and lower latency (30x). Speech quality practitioners can therefore iterate much faster, deploy the system on resource-limited hardware, and, overall, the proposed system contributes to sustainable machine learning. The paper also concludes that framewise embeddings outperform utterance-level embeddings and that multi-task training with acoustic conditions modeling does not degrade speech quality prediction while providing better interpretation.
\end{abstract}

\section{Introduction}
\label{sec:intro}
In communication systems, speech transmission can be easily perturbed by a variety of factors such as environmental conditions (e.g., noisy background), or hardware and software problems (e.g., codec or network impairments). Consequently, providing a precise estimate of speech quality is of valuable interest for various applications, including telephony, speech recognition and synthesis systems, and hearing aids.

Over the past decade, a substantial body of research in speech quality assessment (SQA) has relied on deep neural network-based frameworks~\cite{microsoft_cnn_1, microsoft_cnn_2, qualitynet, nisqa_1, nisqa_2, manocha2021a}. Several objective metrics have been developed for SQA~\cite{rix2001, manocha2021b}, but subjective evaluation based on the mean opinion score (MOS) of human listeners remains the gold standard~\cite{loizou2011}. Producing robust systems using the latter remains challenging due to the difficulty of obtaining large amounts of human-labeled data. To address this issue, a large corpus of 86k speech files was recently released as part of the ConferencingSpeech 2022 challenge~\cite{yi22b_interspeech}. One of the main observations from the challenge's results was that pre-trained general-purpose speech representations were highly effective for SQA. For instance, the top performing approach~\cite{tamm22_interspeech} relied on XLS-R~\cite{xls-r}, a cross-lingually pre-trained version of wav2vec 2.0~\cite{wav2vec2}. Although highly effective, wav2vec 2.0 remains a large model (315M parameters) and may be impractical to use in many applications, e.g., in embedded devices due to its high memory footprint~\cite{gondi2022}.

To this end, in this paper, we explored whether BYOL-S~\cite{scheidwasser2022, elbanna2022byol}, a lightweight speech-specific representation (5M parameters) that achieved competitive performance in the HEAR benchmark~\cite{HEAR_benchmark}, could be used for non-intrusive SQA without loss of performance compared to~\cite{tamm22_interspeech}. To establish a comprehensive comparison, we explored two aspects that should characterize a robust SQA framework. First, in the same vein as the HEAR benchmark, we compared the generation of utterance-level (one embedding per audio file) and framewise (one embedding per audio frame) embeddings on a multilingual benchmark for SQA. Second, we re-trained framewise models using the MOSRA joint learning framework~\cite{elhajal22_interspeech} to assess whether learning room acoustics information can further improve speech quality assessment. The results show that:
\begin{itemize}
    \item BYOL-S rivals the performance of XLS-R with $\sim$40-60$\times$ less parameters, $\sim$10-15$\times$ less memory, $\sim$30$\times$ less latency, and $\sim$100$\times$ less FLOPS. 
    \item Framewise embeddings are more suited for SQA, especially for BYOL-S.
    \item Models trained to jointly predict room acoustics descriptors and speech quality outperformed baselines for room acoustics characterization. These descriptors provide an additional layer of interpretation to speech quality estimations.
\end{itemize}

\section{Methods}
\label{sec:methods}
\begin{table*}[t!]
\centering
\caption{MOS evaluation corpus. m: male, f: female.}
\begin{tabular}{cccccc}
\toprule
Dataset & Languages & Samp. freq. & Speakers & Samples & Av. duration \\
& & [kHz] & & & [s] \\
\midrule
NISQA\_TEST\_FOR~\cite{mittag21_interspeech} & English (Australian) & 48 & 40m, 40f & 240 & 8.5 \\
NISQA\_TEST\_LIVETALK~\cite{mittag21_interspeech} & German & 48 & 4m, 4f & 232 & 9.3 \\
NISQA\_TEST\_P501~\cite{mittag21_interspeech} & English (British) & 48 & 2m, 2f & 240 & 7.9\\
ITU P.Sup23-exp3o,3a,3c,3d \cite{psup23} & French, Italian, Japanese, English  & 8 & 8m, 8f & 2792 & 8.0\\
TCD-VoIP~\cite{harte2015tcd} & English with various accents & 48 & 23 & 800 & 8.1\\
Logi-ReverbSpeechQuality~\cite{nessler21_interspeech} & English with various accents & 16 & 845 & 4032 & 6.0\\
URAE\_TEST~\cite{URAE21} & English and 11 other languages & 16 & 100 & 9400 & 8.0\\
\bottomrule
\end{tabular}
\label{tab:mosratest}
\end{table*}
All models presented in this work comprise two main stages. First, a pre-trained and general-purpose speech model that extracts utterance-level or framewise embeddings from raw audio samples (Sec.~\ref{sec:pretrained}). Second, a downstream network that is trained to assess speech quality on a diverse multilingual corpus (Sec.~\ref{sec:embeddings}). The resulting models were evaluated both in terms of performance (SQA accuracy) and efficiency (i.e., model capacity and speed).

\subsection{Pre-trained feature extractors}\label{sec:pretrained}

\noindent\textbf{XLS-R} - XLS-R~\cite{xls-r} is a cross-lingual adaptation of wav2vec 2.0~\cite{wav2vec2}, a state-of-the-art speech representation. The model was pre-trained in self-supervised fashion on a large corpus of approximately 436k hours of data spanning 128 languages. Using XLS-R as a feature extractor produced the best performance in the ConferencingSpeech 2022 challenge~\cite{yi22b_interspeech} for non-intrusive SQA.

\noindent\textbf{Hybrid BYOL-S/CvT} \cite{elbanna22_interspeech} - BYOL-S is a speech-specific adaptation of BYOL-A~\cite{niizumi2021byol}, a self-supervised, general-purpose, audio representation derived from BYOL (Bootstrap your own latent)~\cite{grill2020}. Both BYOL-S and BYOL-A rely on the comparison of two augmented views of the same input log-mel spectrogram. To do so, the two augmented spectrograms are first fed into an online network and a target network, respectively. Both networks comprise a CNN-based encoder and a projection head with similar architectures. Subsequently, the online network is trained to predict the output projections from the target network using an additional predictor head module. While BYOL-A was pre-trained on the entire AudioSet~\cite{gemmeke2017audio}, BYOL-S was pre-trained on the speech subset of AudioSet~\cite{scheidwasser2022}. Here, we use a hybrid version of BYOL-S, proposed in~\cite{elbanna2022byol}, with a lightweight Transformer-derived encoder based on the Convolutional vision Transformer (CvT) \cite{wu2021}. This system produced competitive performance against the HEAR benchmark with only 5M parameters~\cite{HEAR_benchmark}. This version (called \textsb{Hybrid BYOL-S/CvT}) features a third network that simply extracts features from openSMILE~\cite{eyben2010opensmile}, a hand-engineered feature set consisting of 6373 acoustic functionals computed over low-level descriptor contours. In this case, the online network is trained to predict both the output from the target network and the openSMILE features, hence the hybrid approach.

\subsection{Speech quality assessment models}\label{sec:embeddings}

\setlength{\tabcolsep}{2.5pt}
\begin{table*}[htbp]
\centering
\caption{Comparison of utterance- and frame-wise embeddings for SQA. For framewise embeddings, only Transformer-based downstream models (transf) are shown here. F: framewise embedding, U: utterance-level embedding. $\uparrow$ (resp. $\downarrow$) indicates higher (resp. lower) is better. The best performing approaches for each task and metric are denoted in bold.}
\begin{tabular}{lrrrr@{\hspace{20pt}}ccc}
\toprule
 & \multicolumn{4}{c}{Model capacity/speed} & \multicolumn{3}{c}{SQA quality} \\
 \midrule
 & \#Params & Memory & Latency & GFLOPS & PCC & RMSE & RMSE\_MAP\\
 & [M] & [MB] & [ms] & & $\uparrow$ & $\downarrow$ & $\downarrow$ \\
\midrule
\textit{\textsb{Baselines:}} \\
MelSpec-MOSRA (F)~\cite{elhajal22_interspeech} & 0.42 & 9.93 & 37.8 & 4.17 & 0.81 & 0.58 & 0.54 \\
XLSR-BiLSTM (F)~\cite{tamm22_interspeech} & 316 & 1478.51 & 649.6 & 214.08 & 0.87 & 0.46 & 0.46 \\
\midrule
\textit{\textsb{Proposed - utterance-level emb.:}} \\
XLSR (U) & 316 & 1507.99 & 620.3 & 213.90 & 0.85 & 0.49 & 0.48\\
Hybrid BYOL-S/CvT (U) & 8.17 & 145.53 & \textsb{21.5} & \textsb{2.32} & 0.74
 & 0.64 & 0.63\\
\midrule
\textit{\textsb{Proposed - framewise emb.:}} \\
XLSR-MOS-transf (F) & 316 & 1476.27 & 624.5 & 214.02 & \textsb{0.88} & \textsb{0.44} & \textsb{0.44} \\
Hybrid BYOL-S/CvT-MOS-transf (F) & \textsb{5.20} & \textsb{96.74} & \textsb{24.0} & \textsb{2.33} & 0.85 & 0.50 & 0.49 \\
\bottomrule
\end{tabular}
\label{tab:results}
\end{table*}
\setlength{\tabcolsep}{6pt}

\subsubsection{Framewise embeddings}

Both XLS-R and Hybrid BYOL-S/CvT were used to produce representations of dimension $T \times D$, where $T$ denotes the number of time frames extracted from a given audio sample and $D$ is the embedding dimension. In the case of Hybrid BYOL-S/CvT, a log-mel spectrogram is computed for each audio sample using a 25 ms window size and a 10 ms hop size. Subsequently, the encoder outputs one 2048-dimensional embedding per 160-ms frame. The embeddings were extracted using the pre-trained weights available online\footnote{\href{https://github.com/GasserElbanna/serab-byols}{https://github.com/GasserElbanna/serab-byols}}. As to XLS-R, we used the 300M parameter version of the model available at\footnote{\href{https://huggingface.co/facebook/wav2vec2-xls-r-300m}{https://huggingface.co/facebook/wav2vec2-xls-r-300m}}, which produces a 1024-dimensional embedding at a 20 ms frame step size.

\begin{figure}[htbp]
    \centering
    \includegraphics[width=\columnwidth]{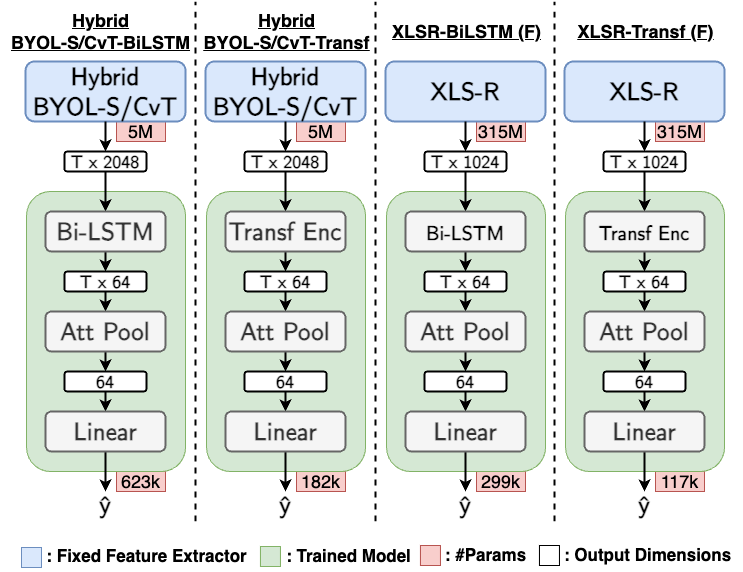}
    \caption{SQA models for framewise embeddings}
    \label{fig:framewise_models}
\end{figure}

\begin{figure}[t!]
    \centering
    \includegraphics[width=0.5\columnwidth]{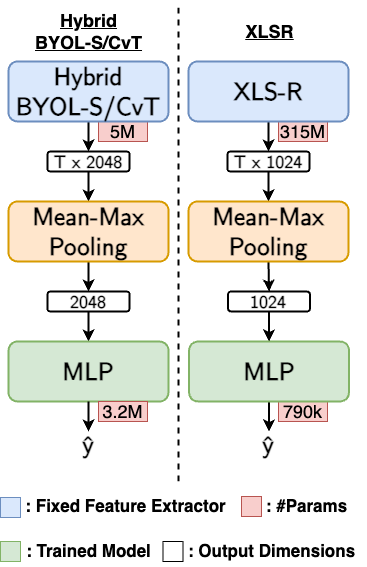}
    \caption{SQA models for utterance-level embeddings}
    \label{fig:scene_models}
\end{figure}

As framewise embedding generation preserves temporal information, we experimented with downstream models based on bidirectional LSTMs (BiLSTMs) and Transformer encoder modules (Fig.~\ref{fig:framewise_models}). Similar to~\cite{tamm22_interspeech}, our first model contained two BiLSTM layers, with each direction having 32 units. On the other hand, the Transformer encoder-based model had two layers with one head and a 64-dimensional hidden layer in the feedforward network. Prior to the Transformer layers, embeddings were first mapped into a 64-dimensional vector using a linear layer. Both downstream models produce a $T \times 64$-dimensional output, which is subsequently temporally reduced using attention pooling~\cite{mittag21_interspeech} before being fed into a linear layer for prediction.

\vspace{-0.25cm}

\subsubsection{Utterance-level embeddings}
Utterance-level embeddings were produced by temporal aggregation of the framewise embeddings using \textit{mean+max} pooling, i.e., $y = \text{max}(x, 1) \oplus \text{mean}(x, 1)$. Thus, for any input audio sample, Hybrid BYOL-S/CvT and XLS-R generate a 1D feature vector of dimension 2048 and 1024, respectively. To estimate speech quality, the embeddings were fed into a multilayer perceptron (MLP) with two ReLU-activated hidden layers, each having a number of neurons equal to half the input dimension (i.e., 1024 for Hybrid BYOL-S/CvT and 512 for XLS-R).

\vspace{-0.3cm}
\section{Experimental setup}
\label{sec:setup}
\subsection{Data}

The training corpus for speech quality assessment comprised the NISQA~\cite{mittag21_interspeech} training sets (11,020 samples, duration: 6-12 s), the Tencent corpus~\cite{yi22b_interspeech} (14,000 samples: 10,000 without reverberation; 4,000 with reverberation), and the PSTN~\cite{pstn} training dataset ($\approx$ 60,000 samples). Validation during training was carried out using the NISQA validation sets and the ITU P. Sup23-exp1 datasets~\cite{psup23}. When applying the MOSRA framework, the URAE data set~\cite{URAE21} was used for room acoustics characterization. The URAE training set contained $\approx$ 80,000 samples with SNR, STI, RT60, C50, C80, and DRR labels, while the validation and test sets comprised 10,000 samples each.

\vspace{-0.3cm}

\subsection{Training procedure}
\label{sec:training_procedure}
Both utterance-level and framewise models were first trained to predict MOS using the above-outlined datasets. Subsequently, as room acoustics information was hypothesized to improve MOS generalization and provide valuable feedback for MOS predictions, we trained framewise models (in a second independent training procedure) using the MOSRA joint training framework~\cite{elhajal22_interspeech}. In this framework, models are trained to simultaneously estimate MOS and various room acoustic characteristics: signal-to-noise ratio (SNR), speech transmission index (STI), reverberation time (T60), direct-to-reverberant ratio (DRR) and speech clarity (C50). The multi-task loss is a weighted sum of each task's mean squared error (MSE) loss.

\subsection{Evaluation}

SQA performance was evaluated against a custom benchmark comprising speech samples recorded in various languages, sampling frequencies and sources of speech degradation. The benchmark, detailed in Table~\ref{tab:mosratest}, includes:
\begin{itemize}
    \item The NISQA test datasets\footnote{\href{https://github.com/gabrielmittag/NISQA/wiki/NISQA-Corpus}{https://github.com/gabrielmittag/NISQA/wiki/NISQA-Corpus}}, which contain various live (e.g., mobile phone, Zoom, Skype, WhatsApp) and simulated (e.g., codecs, packet-loss, background noise) speech quality distortions
    \item The internal ReverbSpeechQuality~\cite{nessler21_interspeech} dataset, which focuses on room acoustics-related effects, such as the reverberation time (T60) of the room impulse responses covering a range up to 1.4 seconds, and background noises matching realistic scenarios like keyboard typing, squeaky chairs, or cafeteria conversations
    \item The TCD-VoIP~\cite{harte2015tcd} dataset, which includes various degradations related to Voice over IP, such as choppy speech, referring to missing samples, competing speakers referring to large-crowd babble, echo effects referring to the receiving unit’s microphone creating a feedback loop and other hardware issues, and clipping effects referring to the amplitude of samples set above the maximum permitted value
    \item The experiment 3 data from the ITU-T P.Sup23~\cite{psup23} coded-speech database that consists of coded and source speech material used in the ITU-T 8 kbit/s codec (Recommendation G.729) characterization tests
\end{itemize} 

Two baselines were considered for both framewise and utterance-level models: the original MelSpec-MOSRA baseline, described in ~\cite{elhajal22_interspeech}, and XLS-R BiLSTM, the system presented in~\cite{tamm22_interspeech} which won the ConferencingSpeech 2022 challenge. Similar to~\cite{yi22b_interspeech}, three metrics were used to assess speech quality: Pearson's correlation coefficient (PCC), and the root mean square error before and after third-order mapping (RMSE and RMSE\_MAP, respectively). Subsequently, following recommendations for fair comparison of model efficiency~\cite{dehghani2022}, four criteria were considered: number of parameters, memory, latency, and FLOPS. All were computed during inference on 30 random audio samples (duration: 5.5-6.5 s) using a single-threaded Intel(R) Core(TM) i7-6850K CPU clocked at 3.60 GHz.

Lastly, for framewise models, we evaluated room acoustics characterization performance after multi-task training using the MOSRA framework. We compared our models against the ``universal room acoustics estimator'' (URAE) model~\cite{URAE21}, and the MelSpec-MOSRA model. The test subset of the URAE dataset is used for evaluation. RMSE was used to quantify the prediction error for each room acoustic characteristic described in Sec.~\ref{sec:training_procedure}.
\section{Results}
\label{sec:results}

Table~\ref{tab:results} compares SQA performance of baseline approaches with models using utterance-level or framewise embeddings as described in Sec.~\ref{sec:methods}. All the proposed models but the utterance-wise model based on Hybrid BYOL-S/CvT embeddings outperformed the MelSpec-MOSRA baseline. Overall, models based on framewise embeddings produced more robust SQA, especially when using Hybrid BYOL-S/CvT for feature extraction. Out of all the models, XLSR-transf produced the best performance across all metrics (PCC, RMSE, and RMSE\_MAP). However, Hybrid BYOL-S/CVT embeddings only produced slightly worse SQA performance than XLS-R embeddings despite a sizeable difference in model capacity ($\sim$40-60$\times$ fewer parameters, $\sim$10-15$\times$ less memory, $\sim$100$\times$ fewer FLOPs, $\sim$30$\times$ less latency). 

Additionally, the results obtained using MOSRA joint training~\cite{elhajal22_interspeech} revealed a modest increase in SQA performance when using Transformer encoders instead of BiLSTMs in the downstream model (Table~\ref{tab:mosresults}). That being said, the multi-lask learning framework did not yield any improvement compared to models trained only to estimate mean opinion scores.

Lastly, the two best models (Hybrid BYOL-S/CvT-transf and XLSR-transf) were benchmarked against the URAE test set for room acoustics characterization (Table~\ref{tab:room_acoustics_results}). Both models outperformed the URAE and MelSpec-MOSRA baselines and achieved similar performance across all five characteristics. Therefore, similar to SQA, we observe that Hybrid BYOL-S/CvT can be used as an efficient feature extractor for room acoustics characterization without loss in performance.

\setlength{\tabcolsep}{3.5pt}
\begin{table}[ht!]
\centering
\caption{MOS task comparison of BYOL-S and XLS-R as feature extractors for models jointly trained using the MOSRA framework.}
\begin{tabular}{lccc}
\toprule
Model & PCC & RMSE & RMSE\_MAP \\
\midrule
Hybrid BYOL-S/CvT-BiLSTM & 0.83 & 0.52 & 0.51 \\
Hybrid BYOL-S/CvT-transf & 0.85 & 0.50 & 0.49 \\
XLSR-transf & \textsb{0.87} & \textsb{0.46} & \textsb{0.45}\\
\bottomrule
\end{tabular}
\label{tab:mosresults}
\end{table}
\setlength{\tabcolsep}{6pt}

\setlength{\tabcolsep}{3.5pt}
\begin{table}[htbp]
\centering
\small
\caption{RMSE comparison of 
 the jointly trained models on the room acoustics characterization tasks.}
\label{tab:room_acoustics_results}
\begin{tabular}{l*{5}c}
\toprule
Model & SNR & STI & DRR & T60 & C50 \\
      & [dB] &    & [dB] & [s] & [dB] \\
\midrule
\textit{\textsb{Baselines:}} \\  
URAE~\cite{URAE21} & 3.36   & 0.080  & 4.60 & 0.46  & 9.60   \\
MelSpec-MOSRA~\cite{elhajal22_interspeech} & 2.63 & 0.073  & 4.21 &  0.45 & 8.84 \\
\midrule
\textit{\textsb{Proposed:}} \\
Hybrid BYOL-S/CvT-transf & \textsb{2.31} & \textsb{0.070} & \textsb{4.00} & 0.43 & 8.40 \\ 
XLSR-transf & 2.42 & 0.071 & 4.01 & \textsb{0.42} & \textsb{8.29} \\ 
\bottomrule
\end{tabular}
\end{table}
\setlength{\tabcolsep}{6pt}
\section{Discussion}
\label{sec:conclusion}

Self-supervised speech representations have proven to be an effective tool for SQA. Indeed, self-supervised frameworks are able to learn knowledge from large data corpora to extract informative and general-purpose features useful for a variety of downstream tasks (e.g., speech recognition, speaker verification, emotion recognition, SQA). Here, we show that Hybrid BYOL-S/CvT~\cite{elbanna22_interspeech} is an efficient speech feature extractor for SQA, achieving competitive results compared to XLS-R while being significantly less resource intensive and much faster. Such a model can be of valuable interest for several applications, namely deployment on memory-limited hardware (e.g., conference room equipment for remote work, hearing aids), or systems requiring real-time SQA (e.g., continuous monitoring of audio transmission during video conferencing or live streaming), and sustainable computing initiatives.

Another notable difference between XLS-R and Hybrid BYOL-S/CvT is granularity. Indeed, XLS-R feature extraction leads to more detailed framewise embeddings, where one framewise embedding corresponds to a 20 ms-frame vs. 160 ms for BYOL-S/CvT. Given the performance improvement when using frame-wise vs. utterance-level embeddings with Hybrid BYOL-S/CvT, future work will examine the impact of granularity on SQA accuracy. In particular, we hypothesize the existence of an optimal frame length, balancing a trade-off between model efficiency and performance.

Regarding the comparison of framewise and utterance-level embeddings, a notable difference was observed when using Hybrid BYOL-S/CvT but not XLS-R for feature extraction. Additionally, the use of Transformer encoding in the downstream model led to slightly superior performance compared to BiLSTM encoding. The fact that Transformer-based models can capture long-term dependencies could allow to better leverage the temporal information from the framewise embeddings.

Furthermore, training our two best models in multi-task fashion to simultaneously predict MOS and room acoustics descriptors did not affect SQA performance. On the other hand, both models significantly outperformed the baseline models in the room acoustics tasks. Thus, using the MOSRA framework can help provide valuable information on SQA predictions, e.g., indicate whether a drop in speech quality is due to increased background noise, late reflections that hamper clarity, or audio distortion.

In conclusion, we show that Hybrid BYOL-S/CvT is an efficient feature extractor for speech quality assessment. This pre-trained model, trained in self-supervised fashion while leveraging information from hand-crafted acoustic features, reached performance levels similar to XLS-R, which topped the ConferencingSpeech 2022 challenge. Future work will strive to assess the performance of hand-crafted feature extraction and additional state-of-the-art frameworks~\cite{serra2021, manocha2021a, yu2021} in order to establish a more comprehensive benchmark. Minimizing the degrees of freedom amongst the different frameworks (e.g., embedding dimension and frame length) should also allow for a more robust analysis of criteria needed for efficient SQA.
%

\ninept
\bibliographystyle{IEEEbib_serab}
\bibliography{refs}

\end{document}